\documentclass[12pt,preprint]{aastex}

\usepackage{natbib}

\slugcomment{Submitted to the Astrophysical Journal}

\begin{document}

\title{Near-Infrared, Kilosecond Variability of the Wisps
 and Jet \\ in the Crab Pulsar Wind Nebula}


\author{A. Melatos and D. Scheltus}
\affil{School of Physics, University of Melbourne,
 Parkville, VIC 3010, Australia}
\email{a.melatos@physics.unimelb.edu.au}
\author{M. T. Whiting}
\affil{School of Physics, University of New South Wales,
 Sydney, NSW 2052, Australia}
\author{S. S. Eikenberry}
\affil{Department of Astronomy, University of Florida, 
 Gainesville, FL 32611, USA}
\author{R. W. Romani}
\affil{Department of Physics, Stanford University,
 Stanford, CA 94305, USA}
\author{F. Rigaut}
\affil{Gemini Observatory, Northern Operations Center,
 670 N. A'ohoku Place, Hilo, HI 96720, USA}
\author{A. Spitkovsky}
\affil{KIPAC, Stanford University, PO Box 20450,
 Stanford, CA 94309, USA}
\author{J. Arons}
\affil{Department of Astronomy, 601 Campbell Hall,
 University of California at Berkeley, 
 Berkeley, CA 94720, USA}
\and
\author{D. J. B. Payne}
\affil{School of Physics, University of Melbourne,
 Parkville, VIC 3010, Australia}


\begin{abstract}
We present a time-lapse sequence of 20 near-infrared
($J$- and $K'$-band)
snapshots of the central $20\arcsec\times 20\arcsec$
of the Crab pulsar wind nebula,
taken at subarcsecond resolution with the Hokupa'a/QUIRC 
adaptive optics camera on the Gemini North Telescope,
and sampled at intervals of 10 minutes and 24 hours.
It is observed that the equatorial wisps and polar knots
in the termination shock of the pulsar wind
appear to fluctuate in brightness on kilosecond time-scales.
Maximum flux variations of
$\pm 24\pm 4$ and $\pm 14\pm4$
per cent relative to the mean (in $1.2\,{\rm ks}$)
are measured for the wisps and knots respectively,
with greatest statistical significance in $J$ band
where the nebula background is less prominent.
The $J$ and $K'$ flux densities imply different near-infrared spectra
for the nonthermal continuum emission from the wisps and 
outermost polar knot (`sprite'),
giving $F_\nu \propto \nu^{-0.56\pm0.12}$ and 
$F_\nu \propto \nu^{-0.21\pm0.13}$ respectively.
The data are compared with existing optical and UV photometry
and applied to constrain theories of the
variability of the wisps (relativistic ion-cyclotron
instability) and knots (relativistic fire hose
instability).
\end{abstract}

\keywords{ISM: individual (Crab Nebula) --- 
 ISM: jets and outflows ---
 pulsars: individual (Crab Pulsar) --- 
 stars: neutron ---
 supernova remnants}


\newpage 

\section{Introduction
 \label{sec:gem1}}
Many young pulsars in supernova remnants
are embedded in synchrotron nebulae,
known as pulsar wind nebulae (PWN),
containing relativistic
electrons and magnetic flux emitted by the central object.
Multiwavelength imaging at subarcsecond resolution
reveals that many PWN share a common morphology,
consisting of 
(i) bipolar jets of unequal brightness, directed along
the (inferred) pulsar spin axis and terminated by
one or more bright knots, 
and (ii) fibrous arcs, or `wisps', that are cylindrically symmetric
about the spin axis, concave or convex with respect to
the pulsar, and associated with an X-ray ring and torus.
Objects known to display this morphology,
likened by some authors to the shape of a crossbow,
include the Crab \citep{hes95,wei00,bie01,hes02,sol03},
Vela \citep{pav01},
G320.4$-$1.2 \citep{gae02}, 
G54.1$+$0.3 \citep{lu02}, 
3C58 \citep{sla02}, 
SNR 0540$-$69.3 \citep{got00},
and G0.9$+$0.1 \citep{gae01},
imaged variously by the
{\em Chandra X-Ray Observatory (CXO)},
{\em Hubble Space Telescope (HST)},
Very Large Telescope (VLT),
and Very Large Array (VLA),
although there are counter-examples too,
e.g.\ G11.2$-$0.3 \citep{kas01}.
The knots and wisps are interpreted, respectively,
as the polar and equatorial zones
of the magnetized, collisionless shock
terminating the pulsar wind
\citep{gal94,hes95,gae02,kom03,spi04}.
The brightness asymmetries are ascribed to Doppler boosting.
Ideas for explaining jet collimation
and the observed jet-torus structure
include magnetic hoop stress in the relativistic wind
upstream from the shock
\citep{beg94,bog01},
the anisotropic energy flux in a force-free, monopole wind
\citep{lyu01,lyu02,kom03,del04},
magnetic hoop stress or hydromagnetic instabilities 
in the subsonic, downstream flow
\citep{beg92,beg98,lyu01,lyu02,mel02,mel04},
interactions with moving ejecta
\citep{pav03},
and the helical structure of a
wave-like, displacement-current-dominated wind
\citep{uso94,hes95,mel96,mel98,mel02}.

The subarcsecond features of PWN are highly variable.
In the Crab PWN,
the optical wisps are seen to change brightness and position 
in time-lapse {\em HST} images every six days,
receding from the pulsar concentrically in a wave-like pattern
at $\approx 0.5c$,
while the optical knots jump around erratically 
on the same time-scale
\citep{hes95,hes02}.
In contemporaneous {\em CXO} images, taken every 22 days,
the X-ray ring inside the torus is resolved into
more than 20 knots that brighten and fade irregularly,
while the X-ray jet transports blobs of material and
bow-wave-shaped structures radially outward at $\approx 0.4c$
\citep{hes02,mor02}.
In time-lapse {\em CXO} images of the Vela PWN,
\citet{pav01,pav03} discovered changes of up to 30 per cent
over several months
in the brightnesses and spectra of the wisps, jet and 
centrifugal ($0.3$--$0.6c$) knots,
coherent bending of the outer jet over 16 days,
and changes in knot brightness over just two days.
The variability has been ascribed to an
ion-cyclotron instability at the shock front \citep{spi04},
to a synchrotron cooling instability \citep{hes02},
and to a nonlinear Kelvin-Helmholtz instability
\citep{beg99}.

In this paper, we report on the first near-infrared,
adaptive-optics observations of the wisps and jet
of the Crab PWN.
The data offer high angular resolution 
(optimum $0.\arcsec19$, average $0.\arcsec23$ in $K'$),
high time resolution 
($0.6\,{\rm ks}$, resolving the light-crossing and
ion-cyclotron time-scales of the narrowest features for
the first time),
and the first feature-specific continuum color
spectra extending 
from near-infrared to ultraviolet wavelengths.
The observations are described in \S\ref{sec:gem2},
the light curves and spectra of individual features
are presented in
\S\ref{sec:gem3} and \S\ref{sec:gem4},
and the results are interpreted physically in \S\ref{sec:gem5}.

\section{Observations
 \label{sec:gem2}}
The center of the Crab PWN was observed on
2002 February 6, 7, and 8
with the Hokupa'a/QUIRC adaptive optics instrument
on the Gemini North Telescope
(proposal GN-2002A-Q-16).
Hokupa'a is a natural guide star, curvature sensing system 
with 36 elements, coupled to a near-infrared camera,
QUIRC, consisting of a $1024\times 1024$ HgCdTe array
with plate scale $0.\arcsec0200$,
dark current $< 0.1$ electrons per second,
and read-out noise 15--30 electrons.
The system is described in detail by \citet{gra98}.
Hokupa'a is capable of locking on to
point sources with $R \lesssim 15\,{\rm mag}$ nominally
and $R \lesssim 17\,{\rm mag}$ in practice.
It is therefore ideal for imaging the Crab PWN,
where there are two suitable guide stars within
$4\arcsec$ of the center of the nebula
(cf.\ isoplanatic radius $\approx 30\arcsec$):
a field star (Star I) at
$\alpha(2000)=05^{\rm h}34^{\rm m}32.2^{\rm s}$ and
$\delta(2000)=+22^\circ00'57''$,
with $R=14.8\pm0.2\,{\rm mag}$,
which we chose to use,
and the Crab pulsar,
with $R=16.3\,{\rm mag}$ pulse-averaged \citep{eik97}.
The wavelengths and bandpasses of the QUIRC $J$ and $K'$ filters 
are $1.25\,\mu{\rm m}/0.17\,\mu{\rm m}$ 
and $2.12\,\mu{\rm m}/0.41\,\mu{\rm m}$ respectively.
We achieved resolutions of 
${\rm FWHM} = 0.\arcsec19$--$0.\arcsec30$ in $K'$ and
${\rm FWHM} = 0.\arcsec24$--$0.\arcsec37$ in $J$
in this sequence of observations,
where FWHM refers to the full-width half-maximum of
the point spread function (PSF).
The seeing fluctuated by up to $\pm0.05\arcsec$ in $K'$
and $\pm0.08\arcsec$ in $J$ over intervals of $0.6\,{\rm ks}$.

We obtained a sequence of $20\times4\times 0.12\,{\rm ks}$ 
exposures spaced in
a four-point, $3\arcsec$ dither pattern,
which were subsequently combined into 20 frames,
as follows:
$5\times K'$ then $3\times J$ (2002 February 6),
$3\times J$ then $5\times K'$ (2002 February 7),
and $4\times K'$ (2002 February 8).
Conditions were excellent on 2002 February 6.
On the following nights,
observations were affected by light cloud and wind.
Data were gathered without interruption during all three nights,
yielding a sampling time of $0.6\,{\rm ks}$ between frames
(after read-out), with two brief exceptions:
a five-point dither was accidentally performed 
on 2002 February 6, 
and two six-point dithers were required on 2002 February 8,
when the telescope lost its guiding.
All features were observed with good signal-to-noise.
In a typical $0.12\,{\rm ks}$ exposure in $K'$,
we accumulated $\approx 4.9\times 10^2$ counts/pixel 
for the pulsar and guide star,
$\approx 4.5\times 10^1$ counts/pixel
for the brightest extended feature 
(the sprite; see \S\ref{sec:gem3}),
and $\approx 2.7\times 10^1$ counts/pixel 
for the faintest extended feature 
(the faint wisp; see \S\ref{sec:gem3}),
{\em after} subtracting the nebula background
($2.9$--$5.1\times 10^3$ counts/pixel, or
 $1.1$--$1.9\,{\rm Jy}\,{\rm arcsec^{-2}}$).

We assembled a data reduction pipeline in 
IRAF\footnote{Image Reduction and Analysis Facility,
Gemini package, v.\ 3.1}
to subtract 
bias, dark, and sky frames and divide by flat fields
in the standard way.
The three bright point sources in the field were excised
from the sky frames with care, to avoid creating false
shadows by oversubtraction.
After trimming a border 20 pixels wide to remove faulty
edge pixels, we median combined each set of four dithered
images, rejecting one high,
then trimmed a border $3\arcsec$ wide to exclude the region 
where the four dithered exposures do not overlap.

An important issue with any adaptive optics observation
is the degree by which the PSF
changes across the field of view.
In this work, we can quantify the effect directly by 
examining the three point sources in the field
(the guide star, the pulsar, and a field star
$6\arcsec$ south of the pulsar, labeled Star II),
which happen to be well separated.
We find that the PSF is
nearly axisymmetric at the southern field star,
elongated in an east-west direction at the pulsar,
and elongated in a northeast-southwest direction at the guide star,
with minor and major axes in the ratio $\approx 0.8$
(although the isophotes are not strictly elliptical).
This essentially precludes accurate photometry of the inner knot,
located $0.\arcsec6$ from the pulsar,
for the reasons set forth in \S\ref{sec:gem3b},
without prejudicing photometry of large, extended
features like the wisps.
Another issue is how the PSF varies as a function of time.
We do not detect any change in the shape of the PSF
at the locations of the three field stars
when comparing isophotes from successive exposures by eye.
However, there is indirect evidence that slight yet rapid 
changes do occur; 
we find that the total flux from the guide star
within an aperture of radius $3.1$ ${\rm FWHM}$
fluctuates by $\pm 6$ ($\pm 4$) per cent in $K'$ ($J$)
over $0.6\,{\rm ks}$ {\em after} background subtraction
(\S\ref{sec:gem3a}),
accompanied by fluctuations of $\pm 9$ ($\pm 4$)
per cent in the southern field star.
These changes feed into the measurement uncertainties
calculated in \S\ref{sec:gem3} and \S\ref{sec:gem4},
as the field stars serve as flux calibrators.

\section{Kilosecond variability
 \label{sec:gem3}}
\begin{figure}
\plotone{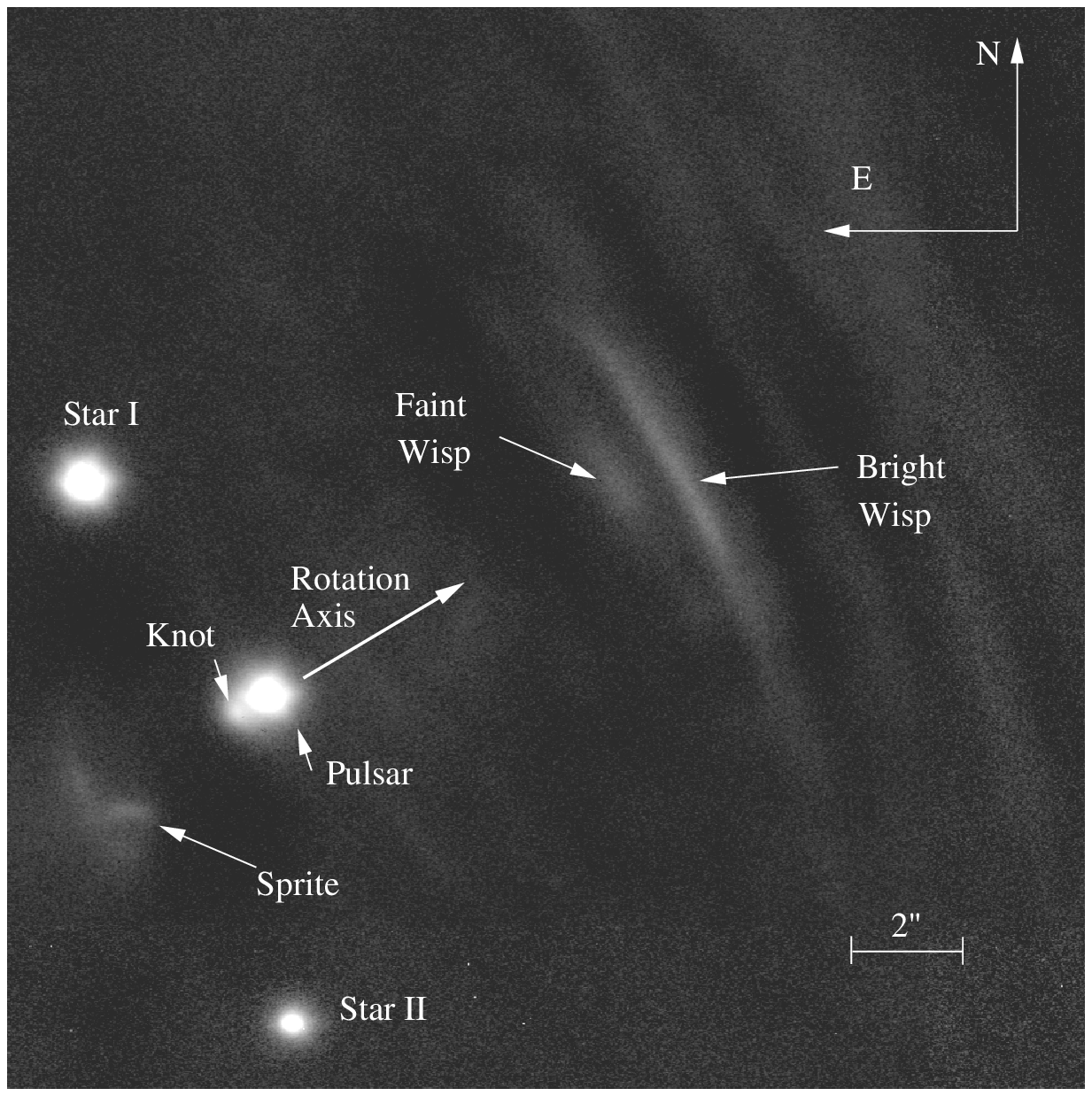}
\caption{
Near-infrared image of the central 
$20\arcsec\times20\arcsec$ region of the Crab PWN,
constructed from four dithered, $0.12\,{\rm ks}$ exposures
in the $K'$ filter, taken on 2002 February 6 with a resolution
of ${\rm FWHM}=0.\arcsec19$.
The principal features of the pulsar wind termination shock
and its environment are labeled,
revealing the `crossbow' morphology characteristic of many PWN:
arc-like, equatorial wisps and jet-like, polar knots
symmetric about the projected rotation axis of the pulsar.
Star I is the Hokupa'a guide star.
Star II is another field star.
\label{fig:gem1}}
\end{figure}

Figure \ref{fig:gem1} shows the center of the Crab PWN
at $0.\arcsec19$ resolution (FWHM) in $K'$,
as it appeared on 2002 February 6.
Its crossbow-like morphology is evident.
The subarcsecond features in the termination shock of the
pulsar wind are cylindrically symmetric about
the projected rotation axis and proper motion
of the pulsar,
determined from {\em HST} astrometry \citep{car99}.
\begin{enumerate}
\item
The {\em wisps} are interpreted as shock structures in the
equatorial plane of the pulsar wind (latitude $\pm10\deg$),
in the neighborhood of the X-ray ring and torus
\citep{hes95,wei00,hes02,mor02,spi04,kom03,del04}.
The faint and bright wisps to the northwest of the pulsar,
labeled in Figure \ref{fig:gem1},
mark ion-driven magnetic compressions
(at the first and second ion turning points)
in the ion-cyclotron model of the shock
\citep{gal94,spi04};
they are probably analogous to the features 
labeled 5 and E by \citet{gae02} 
in another PWN, G320.4$-$1.2.
The position and brightness of these two wisps,
several less prominent wisps, 
and their fibrous substructure are known to
change on a time-scale as short as six days in the
optical \citep{hes02} and 22 days in X-rays
\citep{hes02,mor02}.
\item
The {\em sprite} can be interpreted as a polar shock,
lying on the rotation axis (colatitude $\pm15\deg$)
at or near the base of the polar X-ray jet
\citep{wei00,hes02,mor02}.
It can also be interpreted as a mid-latitude arch shock
between the polar outflow and equatorial
backflow in a pressure-confined, split-monopole nebula,
situated at the tangent point of the line of sight
(and therefore Doppler boosted)
\citep{kom03,del04}.
Its shape, doughnut-like with a central rod (\S\ref{sec:gem3b}),
changes irregularly 
on the same time-scale as the wisps;
in {\em CXO} images, the sprite appears to be the launching point
for blobs and
`bow waves' ejected along the X-ray jet
\citep{hes02,mor02},
although these may also be unstable motions in the
vicinity of the mid-latitude arch shock
\citep{kom03}.
\item
The {\em inner knot} is a barely resolved,
flattened (\S\ref{sec:gem3c}) structure partly obscured
in Figure \ref{fig:gem1} by the pulsar PSF.
It too can be interpreted as a polar feature,
lying $\approx 5$ times closer to the pulsar than the sprite.
If the sprite marks the polar termination shock,
the inner knot sits in the unshocked pulsar wind,
and its physical origin is unknown
\citep{hes95,mel98,mel02}.
If the sprite is a mid-latitude arch shock, 
the inner knot originates from a part of the arch shock
nearer the base of the polar jet,
which is pushed inward (relative to the wisps) because the energy flux
in a split-monopole wind is lower at the poles than at the equator
\citep{kom03}.
\item
A conical {\em halo} is visible at intermediate latitudes,
midway between the pulsar and faint wisp
in Figure \ref{fig:gem1},
the near-infrared counterpart of an optical feature
noted in {\em HST} data by \citet{hes95}. 
We do not discuss the halo further in this paper,
as it is too faint for accurate near-infrared photometry.
\end{enumerate}

In this section, we examine the variability of four features ---
the bright wisp, faint wisp, sprite, and inner knot ---
in the near-infrared 
over time-scales as short as $0.6\,{\rm ks}$,
extending previous studies of the Crab PWN with
{\em HST} (sampling time six days)
and {\em CXO} (sampling time 22 days)
\citep{hes02,mor02}.
The first (and most challenging) step,
subtracting the time-dependent nebula background,
is discussed in \S\ref{sec:gem3a}.
Light curves of the features are presented in
\S\ref{sec:gem3b} and \S\ref{sec:gem3c}.

\subsection{Nebula background
 \label{sec:gem3a}}
It is difficult to characterize and hence subtract
the background in Figure \ref{fig:gem1},
because there is no unique way to disentangle
the contributions from the nebula and sky,
given that the nebula is ubiquitous, nonuniform,
and time-dependent.
The surface brightness $I_{\rm o}$ observed
in any pixel is the sum of flux from
the nebula background ($I_{\rm n}$)
and any feature ($I_{\rm f}$) occupying that pixel,
with
$I_{\rm o} =
 (I_{\rm n} + I_{\rm f}) \eta_{\rm a} + I_{\rm a}$,
where $\eta_{\rm a}$ and $I_{\rm a}$ denote the
atmospheric absorption coefficient
and sky brightness respectively.
There are two problems in extracting $I_{\rm f}$
from $I_{\rm o}$.
First,
there are no pixels empty of both features and nebula,
so we cannot measure $I_{\rm a}$ directly.
In our analysis, we assume that
$\eta_{\rm a}$ and $I_{\rm a}$
are uniform across the field of view,
but both parameters vary markedly from exposure
to exposure, as quantified below.
Second,
$I_{\rm n}$ fluctuates stochastically from pixel to pixel,
so we cannot estimate $I_{\rm n}$ behind a feature 
by interpolating $I_{\rm n}$ directly
from neighboring, feature-free pixels.

To overcome these problems,
we average the observed brightness of all
feature-free pixels ($I_{\rm o}'$) in the field to obtain
$I_{\rm n} \eta_{\rm a} + I_{\rm a}
 = \langle I_{\rm o}' \rangle$
and hence, approximately,
$I_{\rm f} \eta_{\rm a} =
 I_{\rm o} - \langle I_{\rm o}' \rangle$
along any line of sight with a feature,
under the assumption (justified below) that there is no
large-scale gradient of nebula brightness across the field.
The uncertainty in this estimate of $I_{\rm f} \eta_{\rm a}$
is given by the width of the $I_{\rm o}'$ distribution,
measured below.
Without extra, exposure-specific information,
it is impossible to determine 
$\eta_{\rm a}$ and $I_{\rm f}$ independently.
However, we are interested here in the variability of
features rather than their absolute brightness.
Consequently, we can normalize the observed brightness 
of any feature, $I_{\rm o,f}$,
to the observed flux $I_{\rm o,g}$ 
of an intrinsically steady point source $I_{\rm g}$
(e.g.\ the guide star or pulsar),
after subtracting the nebula background,
to obtain
$I_{\rm f}/I_{\rm g} =
 (I_{\rm o,f} - \langle I_{\rm o}' \rangle ) /
 (I_{\rm o,g} - \langle I_{\rm o}' \rangle ) $.
If it were necessary to determine $I_{\rm f}$ absolutely, 
we would need to measure $I_{\rm g}$ independently,
e.g.\ from $J$ and $K'$ photometry of the
the Crab pulsar \citep{eik97}.

\begin{figure}
\includegraphics[angle=90,height=6cm]{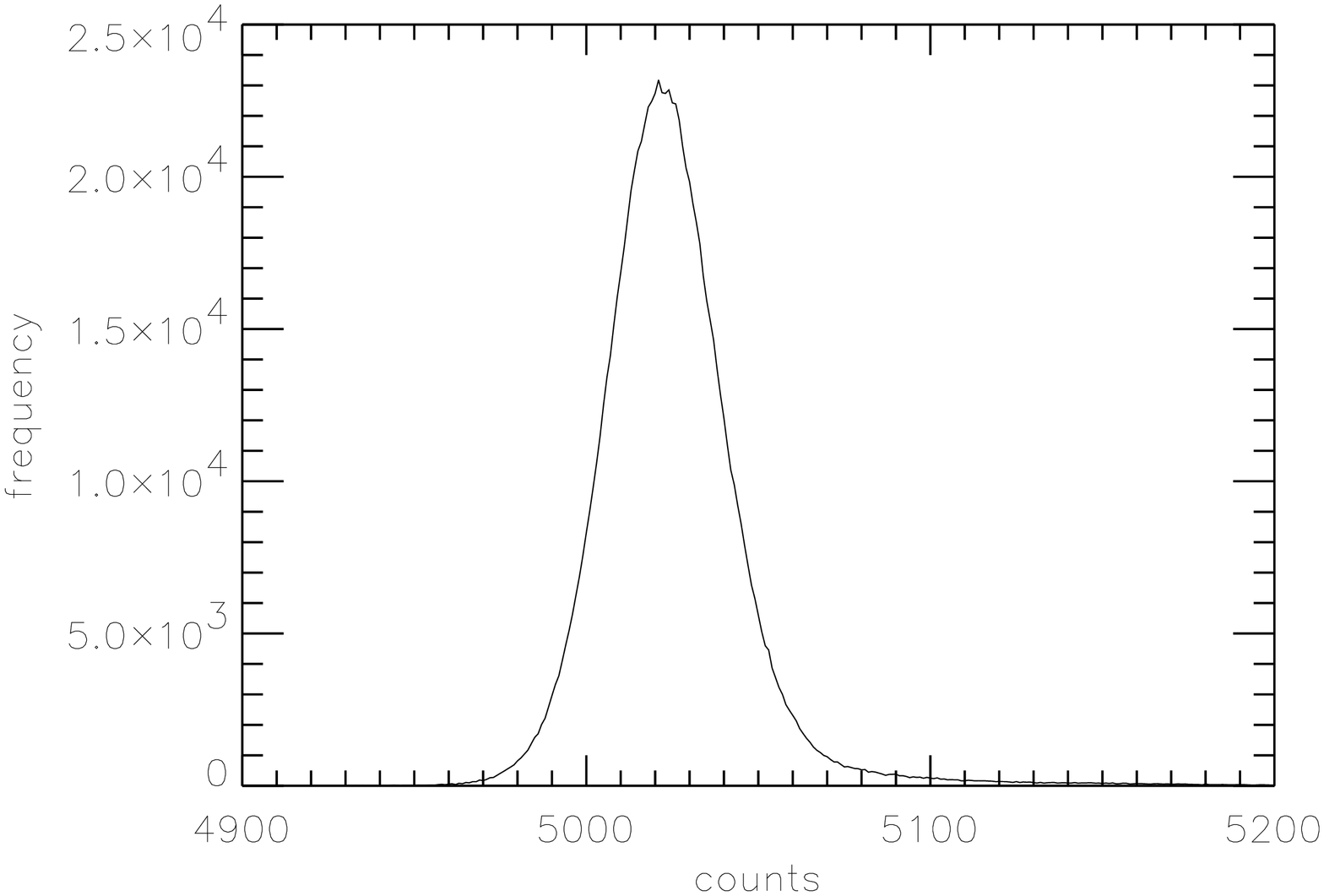}
\includegraphics[angle=90,height=6cm]{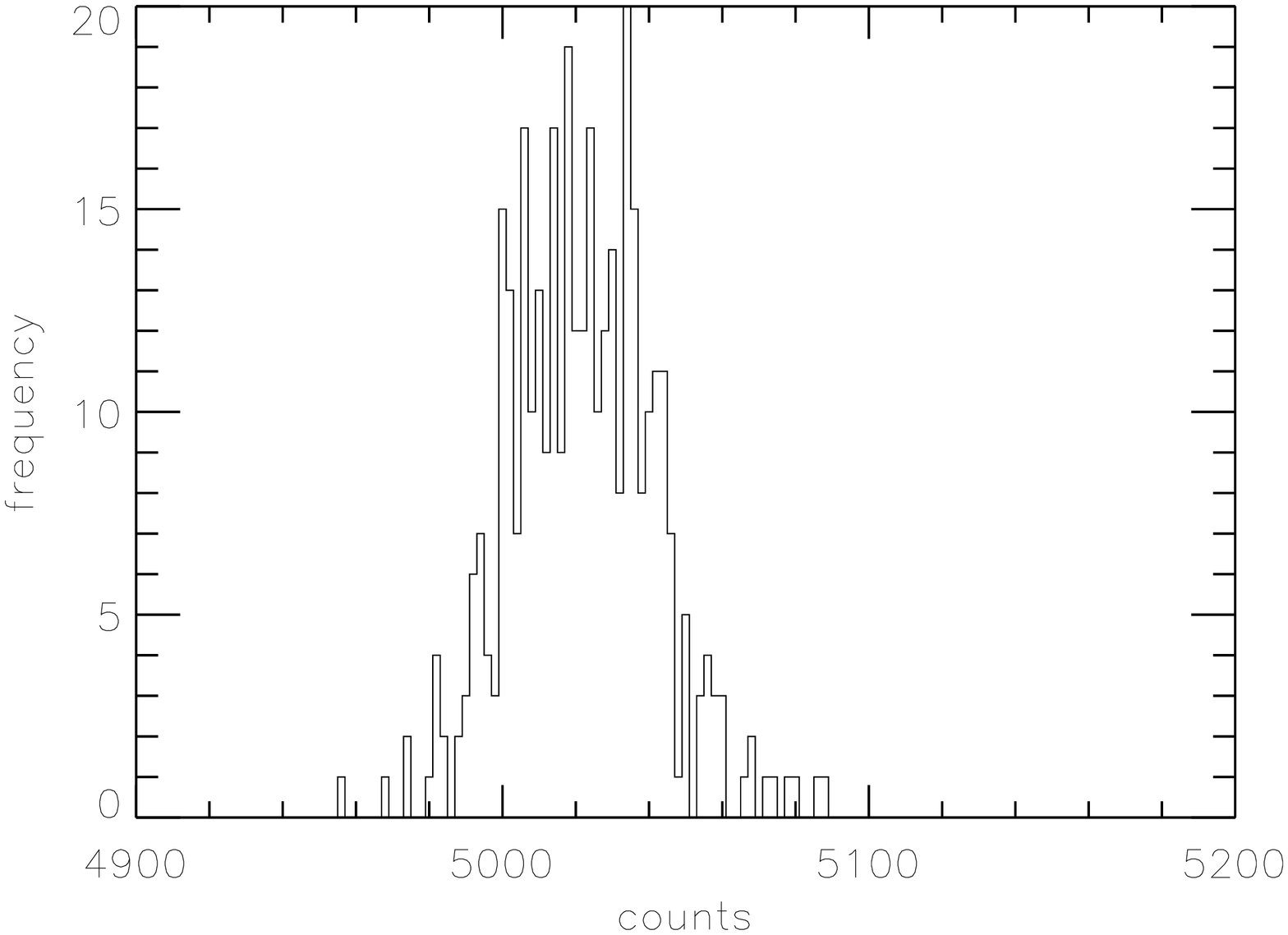}
\caption{
Statistics of the nebula background.
$a$. {\em Left.}
Frequency histogram of pixel counts in a typical $K'$ image
constructed from $4\times0.12{\rm ks}$ dithered exposures
on 2002 February 6,
{\em excluding} point sources and extended features.
Raw counts (solid curve) and counts averaged over
$50\times50$ pixel blocks (dashed curve) are shown;
the curves overlap and are indistinguishable to the eye.
$b$. {\em Right.}
Frequency histogram of pixel counts in an annulus,
radius $6.0{\rm FWHM}$ and one pixel thick,
centered on the guide star (Star I).
\label{fig:gem2}}
\end{figure}

How accurate is the above approach?
A histogram of pixel counts for a single $K'$ frame,
excluding point sources and extended features,
is presented in Figure \ref{fig:gem2}$a$
(solid curve).
The mean, median, and standard deviation ($\delta I$) 
of the distribution in Figure \ref{fig:gem2}$a$
are 5033, 5024, and 13.2 counts respectively.
The distribution is not Gaussian,
cutting off sharply at $\pm 3.5 \delta I$,
and it is narrower than Poisson
($\delta I = 
 0.28 \langle I_{\rm o}' \rangle^{1/2}$)
because $I_{\rm n}$ is correlated in neighboring pixels.
For our $20$ images,
we find $\delta I$ in the range $10$--$30$ counts,
while the median varies markedly in the range
$2.9$--$5.1$ kcounts.
These statistics are corroborated by the dashed curve
in Figure \ref{fig:gem2}$a$,
a histogram of pixel counts averaged over
$50\times 50$ pixel blocks
(chosen to roughly match the dimensions of
the knot-like features of interest).
By inspecting the image directly,
we isolate the blocks that appear to be empty of features,
obtaining a mean, median and standard deviation of 
$5033$, $5024$, and $13.4$
counts per block, in close agreement with the solid curve
(they are indistinguishable to the eye).
We also verify by inspection that there is no large-scale
gradient in counts per block across the field of view,
confirming that $\eta_{\rm a}$ and $I_{\rm a}$ are uniform
within the statistical uncertainty $\delta I$.
Finally, in Figure \ref{fig:gem2}$b$, we present the
histogram of pixel counts in an annular aperture
of radius $6.0$ ${\rm FWHM}$ centered on the guide star.
Annular and polygonal apertures are provided for 
background measurements in the Gemini IRAF software
and are employed in \S\ref{sec:gem3b} and \S\ref{sec:gem3c}.
The statistics are consistent with
Figure \ref{fig:gem2}$a$.
We find that the median background in the annulus
differs by at most $8$ counts from the median of the field
in all $20$ images,
well within the standard deviation $\delta I$,
and is arguably a more accurate estimate of the
background locally.

We estimate the uncertainty in our fluxes as follows.
The absolute uncertainty in $I_{\rm f} \eta_{\rm a}$,
the total flux minus the background,
is given by $(e_1^2 + e_2^2 + e_3^2)^{1/2}$,
where $e_1$ is the square root of the counts
after background subtraction
(corrected for the ADU-photon ratio),
$e_2$ equals $N_{\rm f}^{1/2} \delta I$ 
($N_{\rm f}$ is the number of pixels in the aperture
enclosing the feature),
and $e_3$ is given by $(N_{\rm f}/N_{\rm s})^{1/2} e_2$
($N_{\rm s}$ is the number of pixels in the aperture
estimating the sky).
A similar uncertainty attaches to the guide star
$I_{\rm g} \eta_{\rm a}$.
Note that $e_1$ represents the Poisson
fluctuation in the intrinsic flux of the feature;
$e_2$ measures the uncertainty 
in the background contribution to the total flux,
characterized by Figure \ref{fig:gem2}$a$ and $\delta I$
(not Poissonian);
and $e_3$ is the uncertainty in the background level
subtracted from the total flux,
corrected for the relative sizes of the sky and
feature apertures.

\subsection{Termination shock: wisps and sprite
 \label{sec:gem3b}}
In this section, 
we examine the variability of the
equatorial and polar zones 
(wisps and sprite)
of the termination shock in the Crab PWN.

\begin{figure}
\plotone{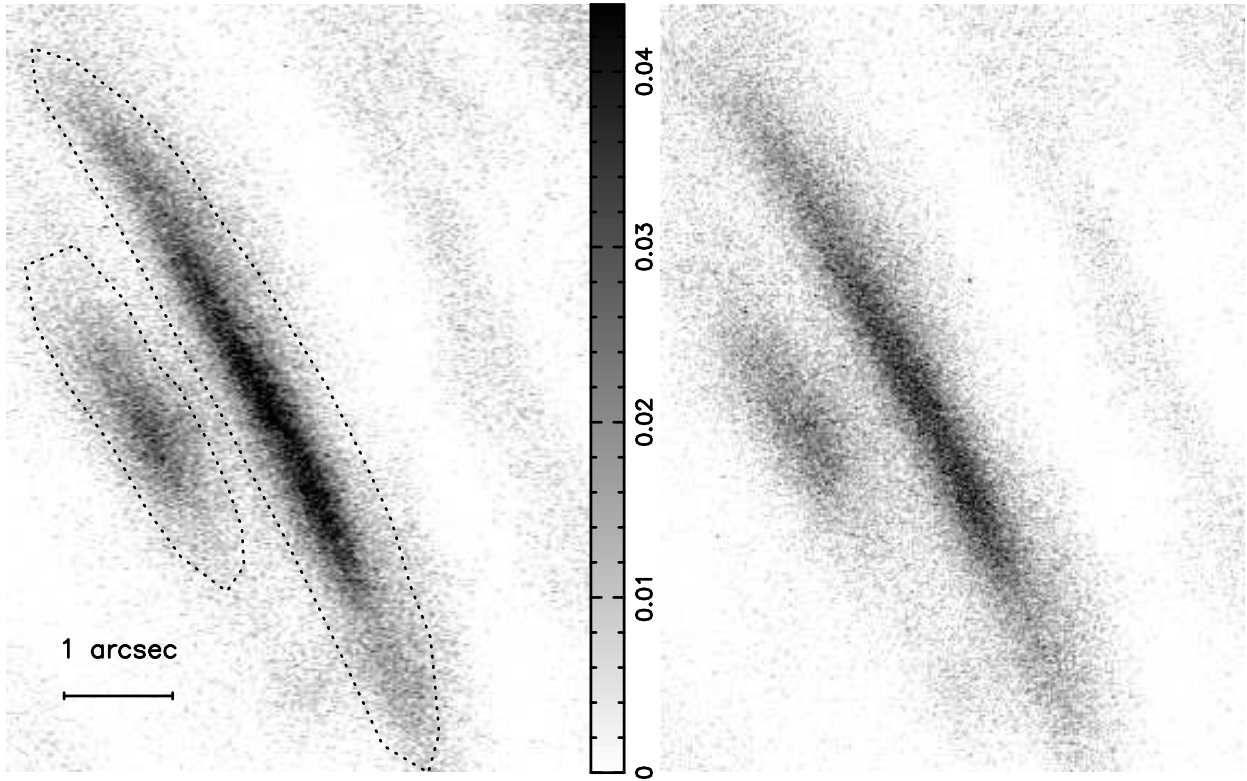}
\caption{
Enlarged $J$-band snapshots of the bright and faint wisps, 
taken on 2002 February 7, separated by $1.2\,{\rm ks}$.
Orientation is as for Figure \ref{fig:gem1}, with north up 
and east left. Each image has been background-subtracted and 
normalized to the flux of Star I. The gray scale is the same 
in both images and is indicated by the central bar. 
The units on the bar indicate the normalized flux 
(defined as in Figure \ref{fig:gem4}) in a $1\arcsec\times 1\arcsec$ box.
The apertures used to measure the total flux of the wisps,
plotted in Figure \ref{fig:gem4}, are drawn as dotted curves. 
The seeing and the guide star flux are the same in the two exposures 
to within $0.\arcsec04$ and $4$ per cent respectively.
\label{fig:gem3}}
\end{figure}

\begin{figure}
\plotone{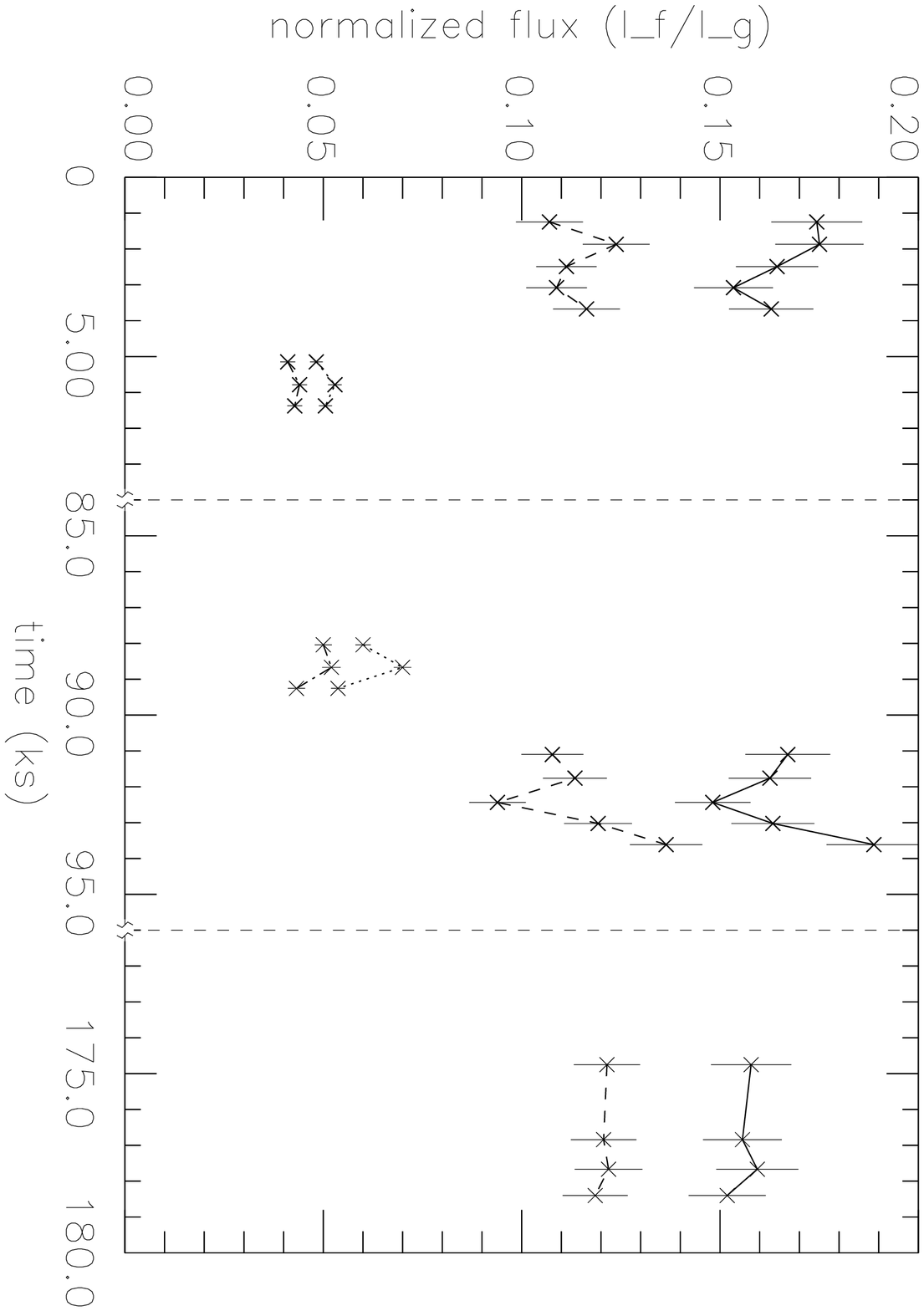}
\caption{
Light curves of the bright and faint wisps,
after nebula subtraction and guide star normalization,
as functions of time (in seconds)
during the three nights of observations
(demarcated by vertical lines).
The four curves represent the
bright wisp in $K'$ (solid),
faint wisp in $K'$ (dashed),
bright wisp in $J$ (dotted),
and
faint wisp in $J$ (dash-dotted).
\label{fig:gem4}}
\end{figure}

Figure \ref{fig:gem3} shows two enlarged images
of the bright and faint wisps in $J$ band,
taken $1.2\,{\rm ks}$ apart on 2002 February 7,
after nebula subtraction and normalization
to the guide star
($I_{\rm f}/I_{\rm g}$; see \S\ref{sec:gem3a}).
The frames should be identical if there is no change in
the intrinsic brightness of the wisps,
yet differences between them are readily apparent
(although the features are not displaced).
To quantify the changes, we use the {\tt polymark}
tool in IRAF to specify a polygonal aperture 
enclosing each wisp, as drawn in Figure \ref{fig:gem3}.
The fluxes enclosed by the apertures,
after nebula subtraction and normalization,
are plotted as functions of time 
in Figure \ref{fig:gem4}.
$J$- and $K'$-band data are both displayed;
uncertainties are calculated according to the recipe
in \S\ref{sec:gem3a}.
For the bright wisp,
we find maximum flux changes (relative to the mean level)
of $\pm 24\pm4$ per cent in $J$
and $\pm 12\pm7$ per cent in $K'$,
occurring in the space of $1.2\,{\rm ks}$ on
2002 February 7,
and smaller fluctuations at other times.
Moreover, the light curves of the bright and faint wisps 
appear correlated to some degree in both filters.
The detection of variability is marginal in $K'$
but more statistically significant in $J$, 
where the nebula background is less prominent.
We find, by experimentation, that the results are 
essentially independent of the choice of aperture,
while the time-dependent PSF has a minimal effect
on the measured flux of the extended features 
(for the guide star,
the effect is included in the measurement uncertainty;
see \S\ref{sec:gem2}).
Nevertheless, new observations --- 
preferably by an independent party 
using a different instrument ---
need to be made
before variability of the wisps on such short time-scales
can confidently be claimed.

\begin{figure}
\plotone{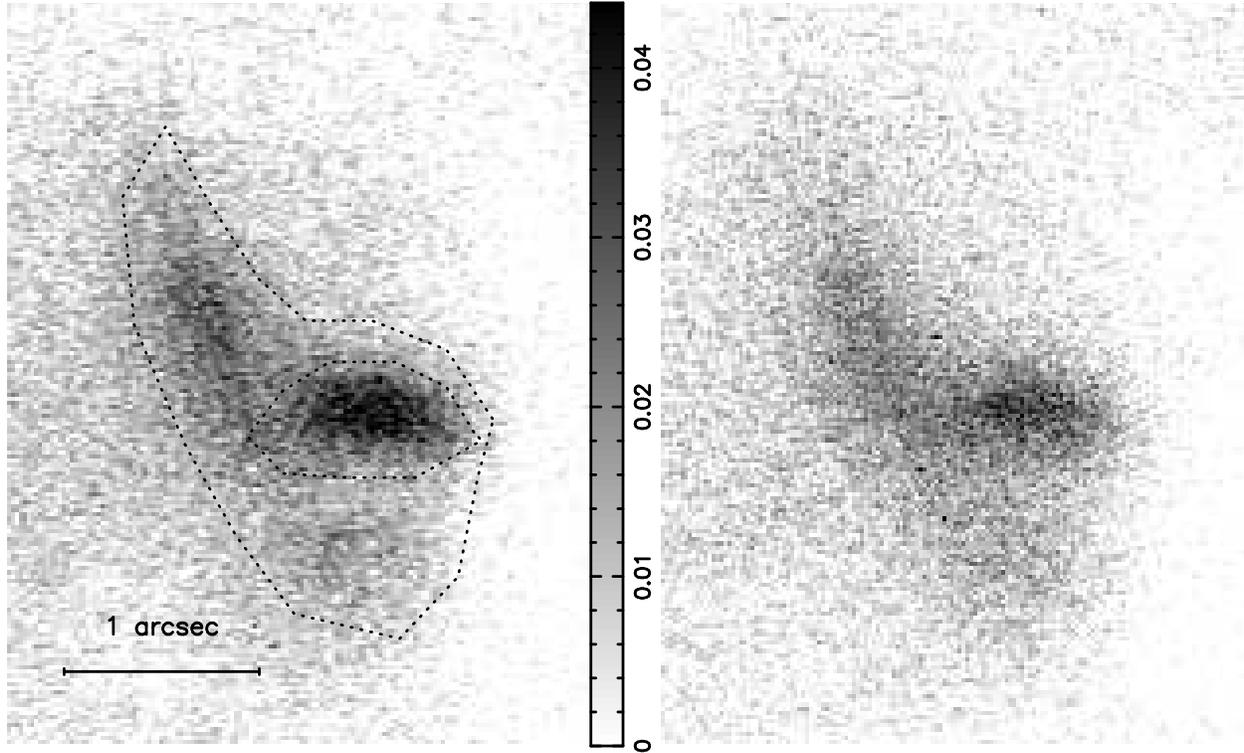}
\caption{
Enlarged $J$-band snapshots of the sprite (the fainter, larger 
feature on the left) and the rod (the brighter, smaller feature 
at center right), 
taken on 2002 February 7, separated by $1.2\,{\rm ks}$.
Orientation is as for Figure \ref{fig:gem1}, with north up 
and east left. Each image has been 
background-subtracted and normalized to the flux of Star I. The gray 
scale is the same in both images and is indicated by the central bar. 
The units on the bar indicate the normalized flux (defined as in
Figure \ref{fig:gem4}) in a $1\arcsec\times 1\arcsec$ box.
The apertures used to measure the total 
fluxes of the sprite plus rod (plotted in Figure \ref{fig:gem6}) 
and the rod alone (quoted in Table \ref{tab:gem1}) 
are drawn as dotted polygons.
The seeing and the guide star flux are the same in the two exposures 
to within $0.\arcsec04$ and $4$ per cent respectively.
\label{fig:gem5}}
\end{figure}

Figure \ref{fig:gem5} shows two enlarged images of
the sprite in $J$ band,
taken on 2002 February 6 and 8.
It is interesting to note its doughnut-like structure,
symmetric about the pulsar's rotation axis,
as well as the short, bent rod emerging from its center,
seen clearly here for the first time and corroborating
the observation by \citet{hes02} that the sprite is often
center-filled (especially in X-rays).
Knot-like features in another PWN, G320.4$-$1.2,
numbered 2 and 3 by \citet{gae02},
may be analogs of the sprite.

\begin{figure}
\plotone{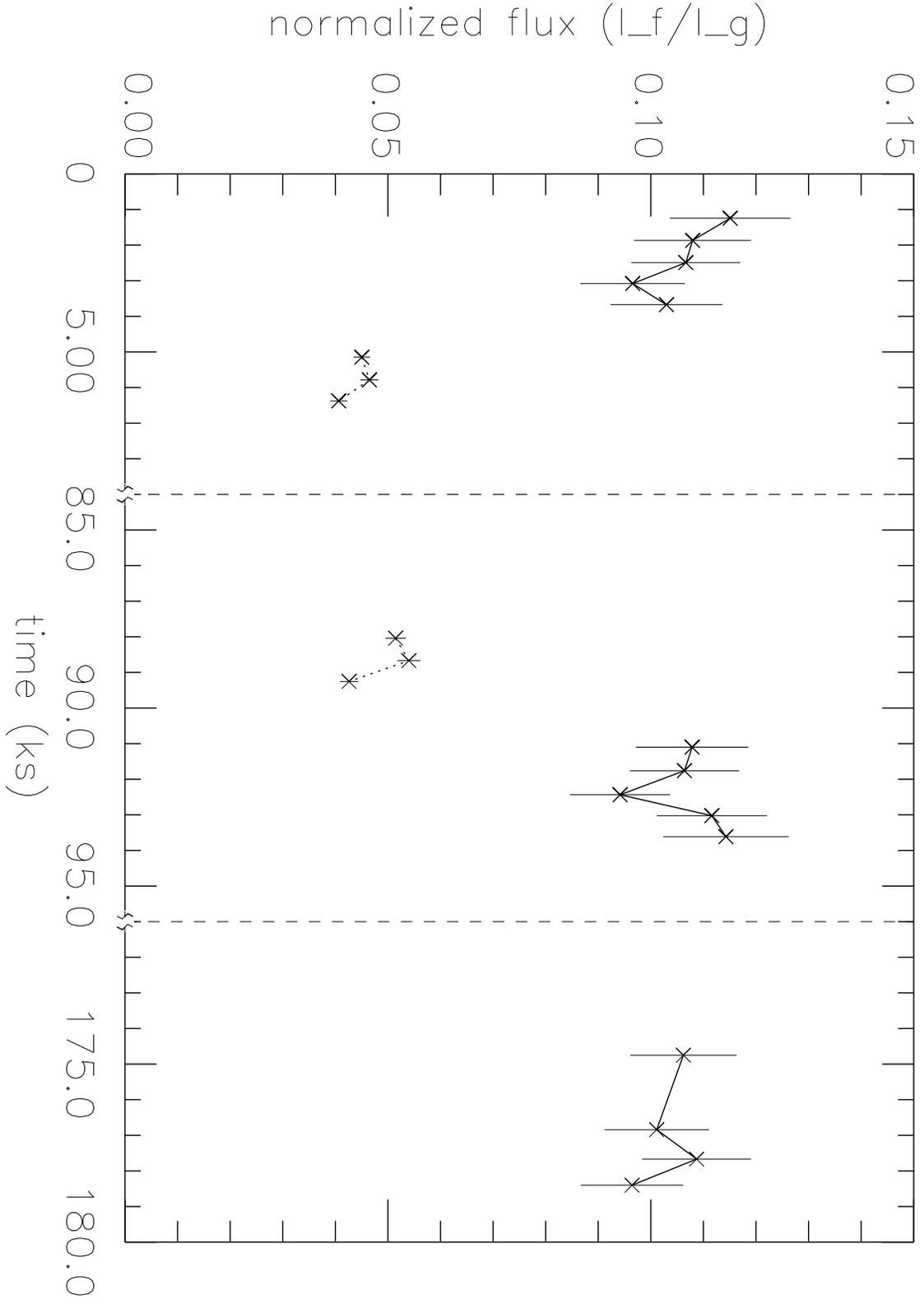}
\caption{
Light curve of the sprite,
after nebula subtraction and guide star normalization,
as a function of time (in seconds)
during the three nights of observations
(demarcated by vertical lines).
The two curves represent $K'$ (solid) and $J$ (dotted)
data respectively.
\label{fig:gem6}}
\end{figure}

In common with the wisps, 
there are visible differences in the brightness
(but not the position) of the sprite in the two images,
even after nebula subtraction and normalization ---
not just the absolute brightness, but also,
more significantly, the brightness contrast between
the rod and doughnut,
which is less likely to be affected by PSF nonuniformity 
and imperfect sky/nebula subtraction.
(The flux of the guide star is
equal to within 4 per cent in the two snapshots.)
To quantify the brightness changes, we define apertures
enclosing the rod and the whole sprite,
and plot the nebula-subtracted, normalized aperture flux
$I_{\rm f}/I_{\rm g}$
versus time in Figure \ref{fig:gem6}.
We measure the following maximum flux changes 
(relative to the mean):
$\pm13\pm4$ per cent (rod, $J$),
$\pm15\pm7$ per cent (rod, $K'$),
$\pm9\pm4$ per cent (whole sprite, $J$),
and
$\pm8\pm7$ per cent (whole sprite, $K'$).
As for the wisps,
the detection of variability is marginal in $K'$
and somewhat more significant in $J$,
especially for the rod,
as seen in Figure \ref{fig:gem5}.

\subsection{Pulsar wind: inner knot
\label{sec:gem3c}}
\begin{figure}
\plotone{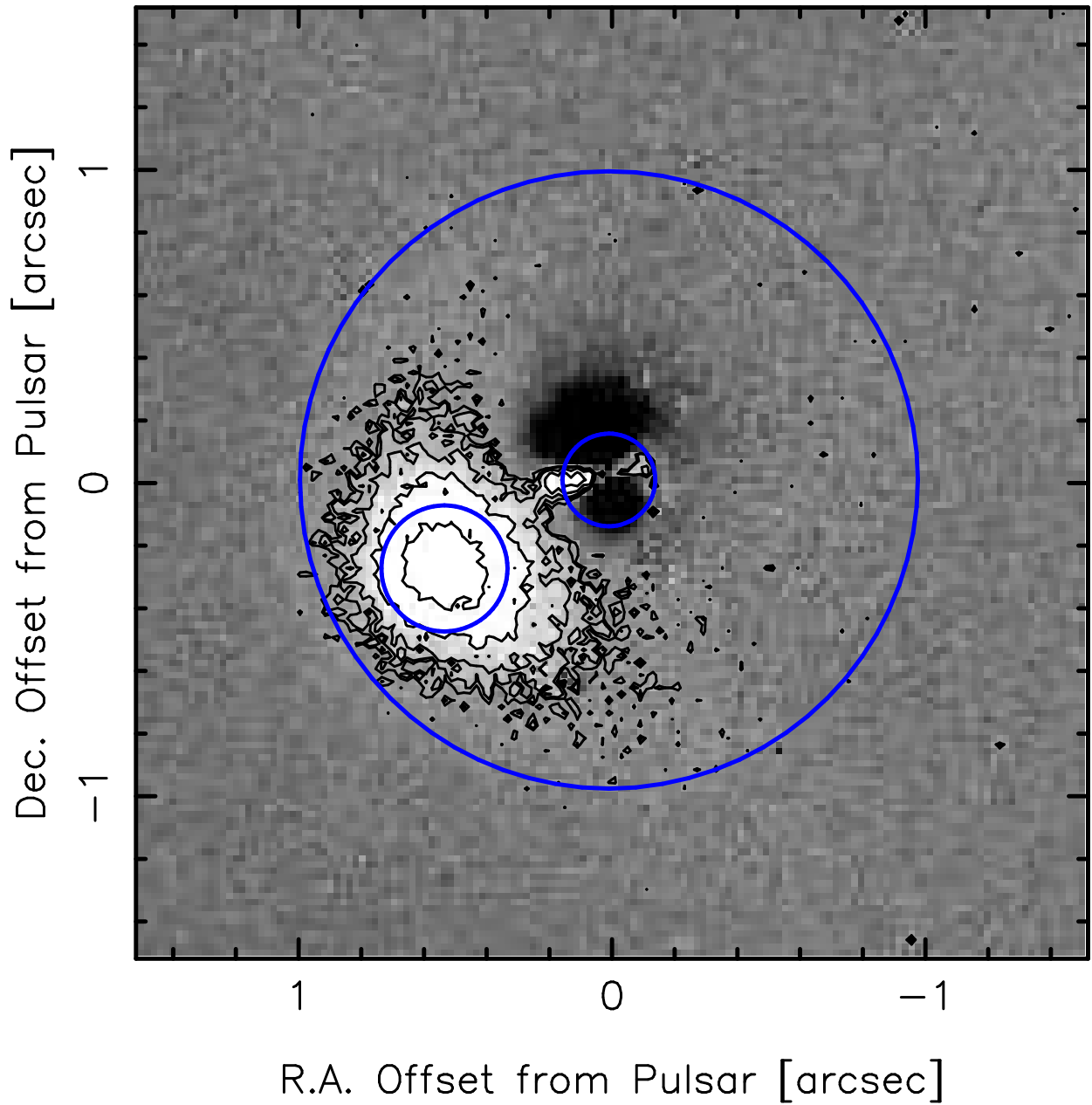}
\caption{
Enlarged $K'$ image of the inner knot after subtraction of 
the pulsar PSF,
plotted as a gray scale image with isophotes 
(solid contours) overlaid.
The subtraction is imperfect because the PSF
template is taken from the guide star, yet the PSF varies
across the field of view (\S\ref{sec:gem2}).
Three apertures used in our photometry of
the inner knot are also shown:
two circles of radius $0.75$ ${\rm FWHM}$
and $5.0$ ${\rm FWHM}$, centered on the pulsar,
and 
a circle of radius 10 pixels, centered on the brightness
centroid of the inner knot.
An annular region of inner radius $6.0$ ${\rm FWHM}$
and one pixel width, also centered on the inner knot
and used to estimate the background
(\S\ref{sec:gem3c}),
lies outside the borders of the image.
\label{fig:gem7}}
\end{figure}

The inner knot, 
discovered by \citet{hes95} in optical {\em HST} data,
is displaced $0.\arcsec65$ from the pulsar
along the axis of symmetry of the PWN,
and is resolved by {\em HST} to be
$\approx 0.\arcsec2$ thick \citep{hes95}.
No counterpart has been detected unambiguously 
at X-ray wavelengths,
although there is a hint of a southeasterly `bump'
protruding from the pulsar in {\em CXO} images,
e.g.\ in Figure 5 of \citet{hes02}.
It is also possible that an analogous feature,
named feature 1,
has been discovered in {\em CXO} images of
another PWN, G320.4$-$1.2 \citep{gae02}.
In Figure \ref{fig:gem7}, we present 
a brightness map of the inner knot,
after subtraction of the PSF.
It is clear, from the isophotes (solid contours) in particular,
that the feature is flattened, not spherical,
although it is hard to discern its shape exactly
because it is barely resolved in our 
highest-resolution $K'$ data
and the PSF subtraction is imperfect.

\begin{figure}
\plotone{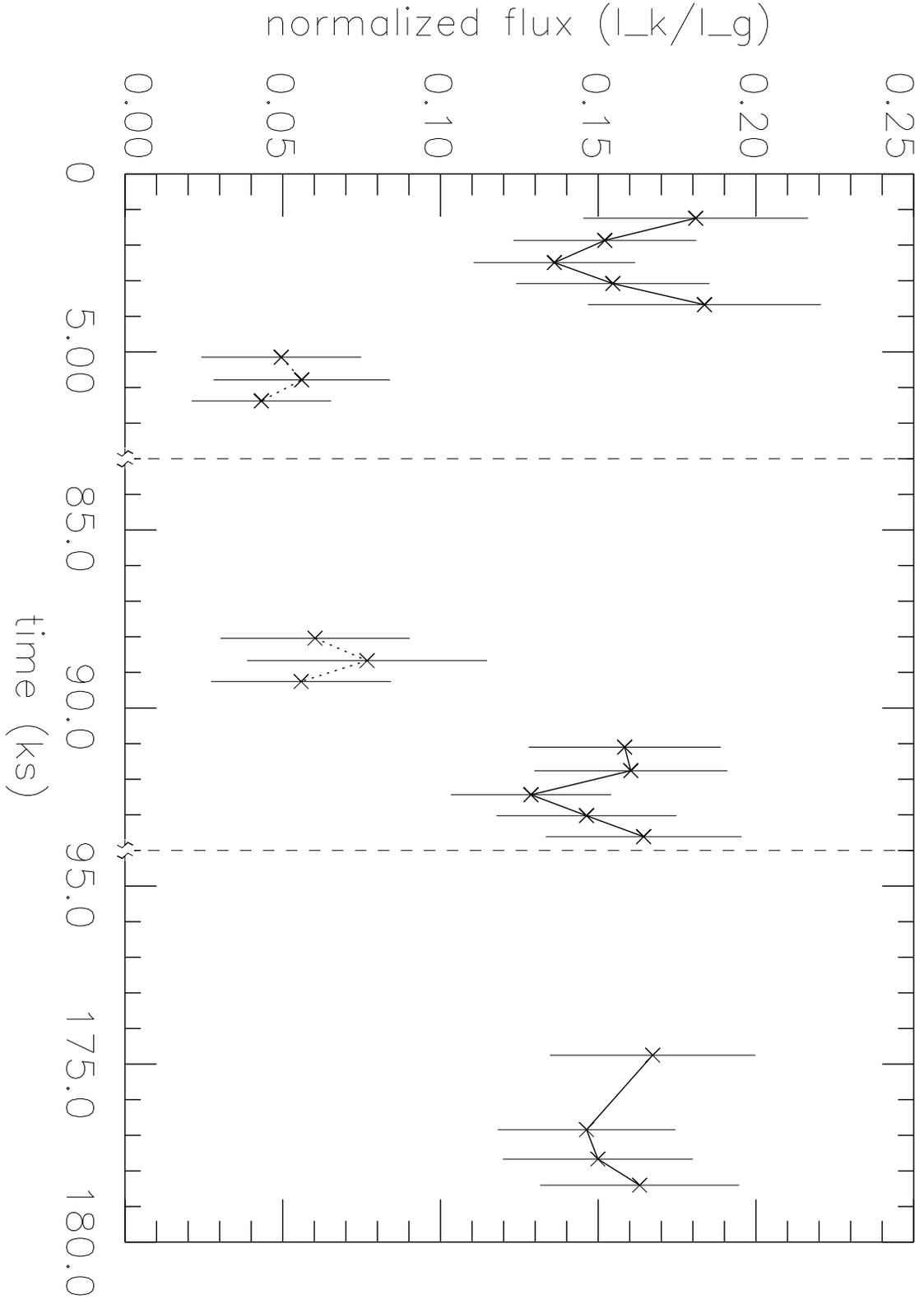}
\caption{
Light curve of the inner knot,
plotted as the flux ratio difference
$I_{\rm k}/I_{\rm g}=
 I_{\rm p}(r_2)/I_{\rm g}(r_2) -
 I_{\rm p}(r_1)/I_{\rm g}(r_1)$
for apertures with $r_1=0.75$ ${\rm FWHM}$ and $r_2=5.0$ ${\rm FWHM}$.
The two curves represent $K'$ (solid) and $J$ (dotted)
data respectively.
\label{fig:gem8}}
\end{figure}

Photometry of the inner knot is complicated by its proximity
to the pulsar,
whose flux contaminates the knot unpredictably
from image to image as the seeing fluctuates.
The standard approach, modeling and subtracting the pulsar PSF,
is attempted in Figure \ref{fig:gem7},
but the result is unreliable; see \S\ref{sec:gem2}.
Faced with these difficulties,
we test for variability of the inner knot
without measuring its brightness directly 
in two complementary ways.
In the first test, we measure the fluxes
$I_{\rm p}(r_1)$ and $I_{\rm p}(r_2)$ 
(after subtracting $\langle I_{\rm o}' \rangle$)
enclosed by two circular apertures centered on the pulsar,
of radii $r_1=0.75$ ${\rm FWHM}$
(including as much PSF as possible but 
excluding most of the knot)
and $r_2=5.0$ ${\rm FWHM}$
(including the PSF and knot). 
The apertures are drawn in Figure \ref{fig:gem7}.
The ratio of these fluxes,
after nebula subtraction, would be the same
as the ratio of the fluxes 
$I_{\rm g}(r_1)$ and $I_{\rm g}(r_2)$ 
enclosed by identical apertures around the guide star
{\em if there were no inner knot},
because the cylindrically averaged PSF is uniform
to within $\pm6$ ($\pm4$) per cent in $K'$ ($J$)  
across the field of view (see \S\ref{sec:gem2}).
Therefore, the flux difference 
$I_{\rm k}/I_{\rm g}=
 I_{\rm p}(r_2)/I_{\rm g}(r_2) -
 I_{\rm p}(r_1)/I_{\rm g}(r_1)$
can be attributed to the presence of the inner knot,
and any change in $I_{\rm k}/I_{\rm g}$
from image to image is evidence that the
inner knot varies intrinsically.
In Figure \ref{fig:gem8}, we plot
$I_{\rm k}/I_{\rm g}$ as a function of time for
the full sequence of observations.
The data are consistent with no variability, 
within the measurement uncertainties.
For example, in the $K'$ band, we find
maximum peak-to-peak changes in $I_{\rm k}/I_{\rm g}$
of $0.045\pm0.060$ over $1.2\,{\rm ks}$
and
$0.054\pm0.063$ over $48$ hours.

\begin{figure}
\plotone{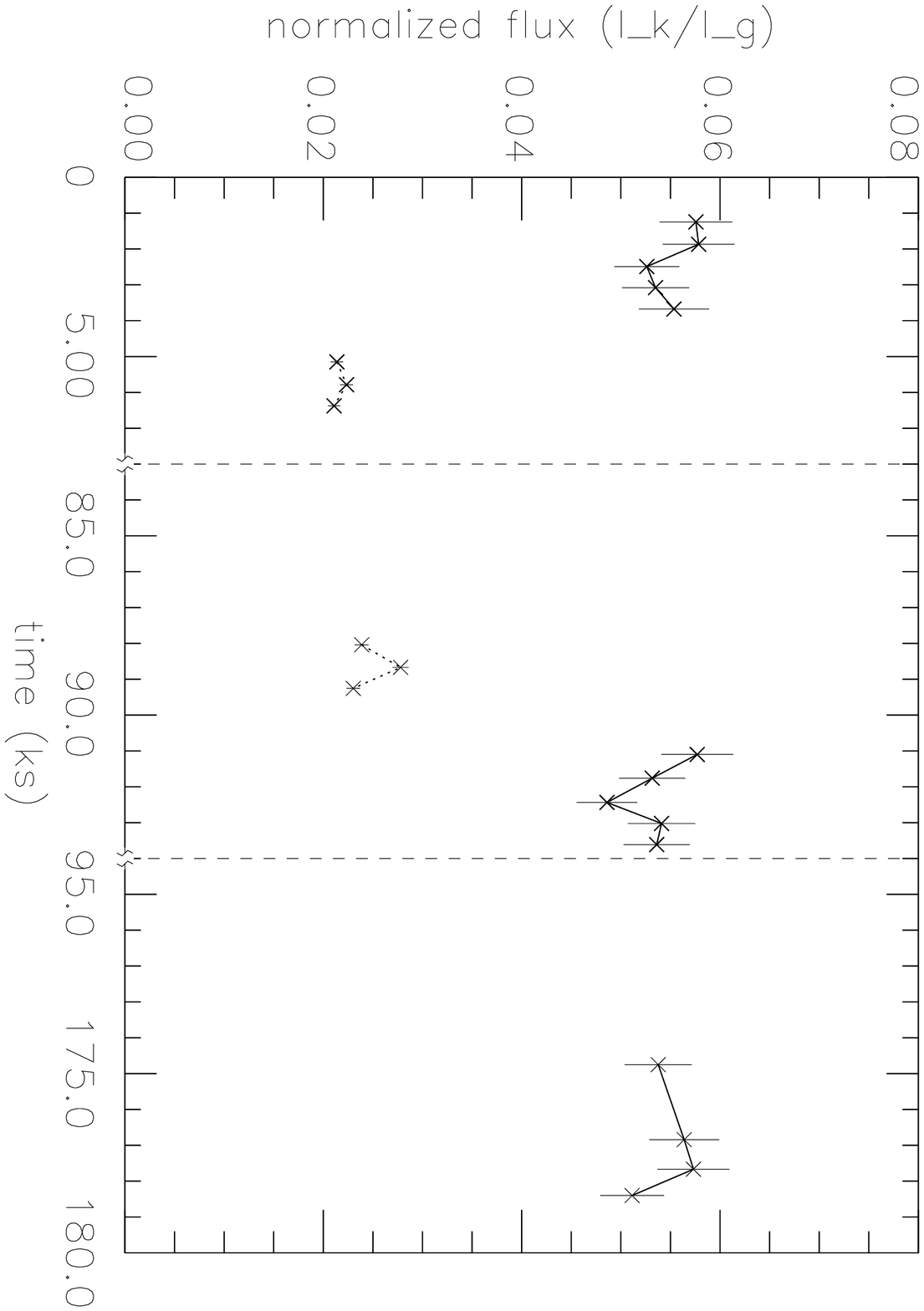}
\caption{
Light curve of the inner knot,
plotted as the flux enclosed by a circular aperture
of radius 10 pixels
(after nebula subtraction and guide star normalization).
This aperture includes some leakage from the pulsar PSF.
The two curves represent $K'$ (solid) and $J$ (dotted)
data respectively.
\label{fig:gem9}}
\end{figure}

A second test provides a consistency check:
we measure directly the flux enclosed by a circular
aperture of radius 10 pixels, centered on the knot.
The flux thus measured includes leakage from the pulsar PSF,
the amount of which varies from
image to image along with the seeing,
as noted above,
but strong intrinsic variations
in knot brightness could still overwhelm this effect.
In Figure \ref{fig:gem9}, we plot the flux in the
10-pixel aperture as a function of time,
after nebula subtraction and guide star normalization.
The result is consistent with Figure \ref{fig:gem8}:
there is no significant detection of variability
within the measurement uncertainties,
with a maximum peak-to-peak change
of $0.011\pm0.007$ over $1.2\,{\rm ks}$
on 2002 February 7.

On the strength of the data presented here,
we are unable to say whether or not the inner knot
is variable on time-scales of $0.6\,{\rm ks}$
to $48$ hours. New observations --- 
preferably by an independent party using a
different instrument ---
are required to settle the issue,
and dedicated PSF calibration frames will be
essential if adaptive optics are used.

\section{Feature-specific color spectra
 \label{sec:gem4}}

\subsection{Near infrared: wisps, sprite, and inner knot
 \label{sec:gem4a}}
In this section, we measure the $J$-to-$K'$ color spectra
of the faint wisp, bright wisp, sprite, rod, and inner knot. 
As these features may vary on time-scales shorter than
the minimum interval between exposures,
we sum the 14 $K'$ images and 6 $J$ images in our data set
to obtain time-averaged spectra.

To the best of our knowledge,
calibrated $J$-to-$K'$ spectra of Stars I and II have not been published.
Therefore, to calibrate the spectra of the various
subarcsecond features in the PWN, 
we are forced to redo the photometry in
\S\ref{sec:gem3} by normalizing nebula-subtracted fluxes 
with respect to the pulsar, 
whose phase-averaged, near-infrared, color spectrum 
was determined by \citet{eik97}
using the Solid State Photomultiplier
on the Multiple Mirror Telescope.
After dereddening, the pulsar's spectrum is fairly flat 
in the near infrared ($F_\nu \propto \nu^{0.36}$);
see the third row of Table \ref{tab:gem1}.
Note that the fluxes in \citet{eik97} include the inner knot
(unresolved from the ground).
By comparing with the total flux of the pulsar plus knot in our images,
we derive a zero magnitude reference for calibrated
aperture photometry of the extended features (e.g.\ wisps, sprite, and rod).

To measure the flux of the knot, we remove the contribution from 
the pulsar by scaling it to the azimuthally averaged PSFs
of Stars I and II,
such that the normalization is the same at a radius of 10 pixels. 
The combined profile of the pulsar and knot is found to match
the stellar PSFs up to a radius of $\approx 20$ pixels 
(to within $0.05\,{\rm mag}$ in $K'$ and $0.03\,{\rm mag}$ in $J$)
but deviates beyond due to the knot excess.
We measure the flux difference between radii of 20 and 60 pixels
(where the profile merges into the background)
and adjust the results in \citet{eik97} to give
the calibrated knot flux.
Note that the differential method of detecting the knot 
($I_{\rm k}/I_{\rm g}$)
employed in \S\ref{sec:gem3c} does not yield a calibrated flux.

Time-averaged $J$- and $K'$-band fluxes are presented 
in Table \ref{tab:gem1}
for the point-like and extended subarcsecond features identified 
in Figure \ref{fig:gem1},
together with the spectral index $\alpha$ of each feature
assuming its flux density scales as $F_\nu \propto \nu^{\alpha}$
(as for synchrotron radiation,
although not necessarily for all nonthermal processes,
e.g.\ synchro-Compton radiation; see \S\ref{sec:gem5}).
The fluxes for the extended features are quoted per unit length,
although it is possible that we are marginally resolving the wisps 
across their width.
The field stars are included for reference,
as their spectra have not been published previously.
Three sources of uncertainty,
summed in quadrature, contribute in Table \ref{tab:gem1}.
First,
the central wavelength of the $K$ filter used
by \citet{eik97}
is $0.08\,\mu{\rm m}$ greater than for the
QUIRC $K'$ filter.
Second, there is scatter in the 20-pixel flux and 20-to-60-pixel offset.
Third,
the extinction corrections are uncertain
($0.63\pm 0.03\,{\rm mag}$ in $J$ and
 $0.81\pm 0.03\,{\rm mag}$ in $K'$).

Several interesting conclusions emerge from these data.
First, the inner knot is clearly the reddest feature in the region,
with $\alpha\approx {-0.8}$.
This is understandable if the inner knot is produced in the 
pulsar wind by a different radiation mechanism than other features.
However, it is surprising if the inner knot is part of the
same arch shock that produces the sprite.
Second, the polar or mid-latitude sprite and rod have
flatter spectra ($\alpha={-0.21\pm0.13}$)
than the equatorial wisps ($\alpha={-0.56\pm0.12}$).
Yet all these features are synchrotron emitting elements
of the same termination shock, albeit at different latitudes.
Third, there is no feature in the region whose spectrum
matches smoothly from the near infrared to X-rays.
The theoretical implications of these results
are considered further in \S\ref{sec:gem5}.

\subsection{Ultraviolet: inner knot
 \label{sec:gem4b}}
We extend the spectrum of the inner knot
in Table \ref{tab:gem1} with an independent measurement 
of the ultraviolet flux of this feature.
\citet{gul98} observed the Crab pulsar with the {\em HST}
Space Telescope Imaging Spectrograph (STIS) NUV-MAMA detector
on 1997 August 7 through the low dispersion G230L grating.  
The observations were made using a $2\arcsec \times 2\arcsec$ 
aperture which included the inner knot,
with part of the exposure ($2 \times 2.4\,{\rm ks}$) in TIME-TAG mode.
We acquired the archival data, barycentered the photons using
standard STIS routines, and extracted a `slit' image using photons 
from $0.25\,{\rm rad}$ of phase spanning the pulse minimum,
thereby gating out the pulsar.
Approximately two per cent of the unpulsed flux remained,
but, after subtracting a scaled version of the on-pulse PSF,
a clear excess was found at the projected position of the
inner knot $\approx 0.3\pm0.2\arcsec$ on one side of the pulsar. 
The intensity profile agrees with that expected from a 
one-dimensional collapse of direct HST images.  
Assuming a flat $\alpha \approx 0$ spectral index over the NUV band 
($160$--$320\,{\rm nm}$),
we find a summed inner knot flux
$(3.3 \pm 0.1) \times 10^{-2}$ times the unpulsed flux of the pulsar.
De-reddening with the best fit value $E(B-V)=0.52$ \citep{sol00}
yields 
$F_\nu = 0.12 \pm 0.01\,{\rm mJy}$
for wavelengths in the range
$0.16$--$0.32\,\mu{\rm m}$.
Note that the ultraviolet flux is consistent with the 
near-infrared spectrum measured in \S\ref{sec:gem4a},
which extrapolates to give $F_\nu = 0.11\,{\rm mJy}$ at
$0.32\,\mu{\rm m}$ for $\alpha=-0.8$.

\section{Discussion
 \label{sec:gem5}}
In this paper, we report on the first near-infrared, 
adaptive-optics observations of the wisps and jet of the
Crab PWN, comprising 20 $J$- and $K'$-band snapshots
taken at $0.\arcsec19$--$0.\arcsec37$ resolution 
with the Hokupa'a/QUIRC camera on the Gemini North Telescope.
The data contain tantalizing --- albeit inconclusive --- 
evidence that subarcsecond features in the termination shock 
of the Crab PWN vary intrinsically in $J$-band brightness 
by $\pm24\pm4$ (wisps) and $\pm14\pm4$ (sprite) per cent
on time-scales as short as $1.2\,{\rm ks}$.
The principal sources of uncertainty
are the nonuniform, unsteady nebula background and PSF.
The data also suggest that the near-infrared spectra 
of polar features in the termination shock are flatter
(e.g.\ sprite, $F_\nu \propto \nu^{-0.21\pm0.13}$)
than the spectra of the equatorial wisps
($F_\nu \propto \nu^{-0.56\pm0.12}$),
except for the steep-spectrum inner knot 
($F_\nu \propto \nu^{-0.8}$),
which may lie in the unshocked pulsar wind.
This result is supported by an independent measurement of the
ultraviolet flux of the inner knot,
obtained by reanalyzing archival, time-tagged, {\em HST} STIS data.

\subsection{Ion cyclotron and fire hose instabilities
 \label{sec:gem5a}}
Why, physically, might the nebula vary on time-scales
as short as $1.2\,{\rm ks}$?
According to one hypothesis,
modeled numerically by \citet{spi04},
the wind contains ions which
drive a relativistic cyclotron instability at the termination shock.
The instability exhibits limit cycle dynamics:
ion bunches and
magnetic compressions are launched downstream periodically
at roughly half the ion-cyclotron period,
$\frac{1}{2} T_{\rm i} = 
 \pi A_{\rm i} m_{\rm p}\gamma_{\rm i} / 
 Z_{\rm i} e B$,
where $A_i$ and $Z_i$ are the
atomic number and charge, $\gamma_i$ is the preshock
Lorentz factor, and $B$ is the postshock magnetic field.
In order to fit the separation of the innermost wisps,
one must take $T_{\rm i} \approx 3.2\times 10^7\,{\rm s}$,
consistent with $\gamma_{\rm i}=7.8\times 10^5$ and
$B=16\,\mu{\rm G}$ at a radial distance $r=0.1\,{\rm pc}$
from the pulsar (for $A_{\rm i} / Z_{\rm i}=1$)
\citep{spi04}.
Faster variability is expected nearer the pulsar,
because the magnetic field in the wind scales
as $B=16(r/0.1\,{\rm pc})^{-1}\,\mu{\rm G}$.
\footnote{Faster variability is also expected at certain
special phases in the month-long ion cycle,
e.g. in the neighborhood of a moving wisp,
where the plasma is stirred up 
(A. Spitkovsky, private communication).}
However, we find
$T_{\rm i}\approx 2\times 10^7\,{\rm s}$
and
$T_{\rm i}\approx 2\times 10^6\,{\rm s}$
at the sprite ($r\approx 0.05\,{\rm pc}$)
and inner knot ($r\approx 0.007\,{\rm pc}$) respectively,
slower than the variability observed.
Relativistic Doppler boosting does not improve the agreement;
the time-scale $\propto \gamma_{\rm i}^{-2}$ 
is unchanged downstream 
($\gamma_{\rm i}\sim 1$) and too short upstream 
($\gamma_{\rm i}\sim 10^6$).
The electron-cyclotron period,
$T_{\rm e}=(m_{\rm e}Z_{\rm i}/m_{\rm p}A_{\rm i})T_{\rm i}$,
does fall in the observed range,
but it is hard to see how to maintain
coherent limit-cycle dynamics in the electrons when they are
randomized rapidly at the shock by ion-driven magnetosonic waves
\citep{gal94,spi04}.
We are therefore inclined to rule out a cyclotron origin
of the observed kilosecond variability in the near infrared.

The argument against a cyclotron origin of the kilosecond
variability assumes that all the energy in the unstable
(compressional) ion-cyclotron-magnetosonic waves 
resides in the fundamental. This need not be so.
The frequency spectrum of the waves is quite flat in
one-dimensional simulations \citep{hos92};
significant power is deposited at high harmonics
(up to orders $\sim m_{\rm p}/m_{\rm e}$)
provided that parametric three-wave decays do not destroy 
the coherence of the waves, a plausible concern in a realistic, 
three-dimensional plasma (J. Arons, private communication).

Another possible mechanism,
applicable especially to the knots in the polar jet,
is the relativistic fire hose instability,
driven by anisotropy of the kinetic pressure
parallel ($P_\parallel$)
and perpendicular ($P_\perp$) to the magnetic field $B$.
\citet{noe69} showed that the growth time $T_{\rm fh}'$ 
in the bulk frame of the jet
(denoted by primes) is given by
\begin{equation}
 \frac{1}{T_{\rm fh}'} = 
 \frac{eB'}{m_{\rm e} \langle\gamma'\rangle}
 \frac{1.3-B'^2/[\mu_0 (P'_\parallel-P'_\perp)]}
  {(1+c^2/v_{\rm A}'^2)^{1/2} + 8}~,
\label{eq:gem1}
\end{equation}
where $\langle\gamma'\rangle$ is the thermal Lorentz factor,
$v_{\rm A}'^2=B'^2/\mu_0\rho'$ is the Alfv\'{e}n speed,
and $\rho'$ is the density of the cold ion background.
As long as the condition 
$P_\parallel'-P_\perp'>0.77B'^2/\mu_0$ 
is met, the minimum growth time is given by
$T_{\rm fh,min}'=0.14 m_{\rm e} \langle\gamma'\rangle/eB'$
for $v_{\rm A}'\gg c$.
Upon Lorentz transforming to the observer's frame,
we obtain 
(i)
$T_{\rm fh,min}=
 0.14 m_{\rm e} \langle\gamma\rangle/eB$
if $B$ is radial in the jet,
or (ii)
$T_{\rm fh,min}=
 0.14 \gamma_{\rm i} m_{\rm e} \langle\gamma\rangle/eB$
if $B$ is helical or toroidal in the jet,
assuming 
$\langle\gamma\rangle \approx
 \gamma_{\rm i}\langle\gamma'\rangle$.
(The validity of the last assumption depends subtly on the 
exact form of the electron distribution.)
Immediately upstream from the termination shock
of the jet, we have $\langle\gamma\rangle = 1$,
$B=2.7\,\mu{\rm G}$, and $\gamma_{\rm i}=7.8\times 10^5$,
implying $T_{\rm fh,min}=2.3\,{\rm ks}$ in scenario (ii).
Immediately downstream from the termination shock,
we have $\langle\gamma\rangle \approx 1\times 10^6$,
$B=8\,\mu{\rm G}$, and $\gamma_{\rm i}=1.1$,
implying $T_{\rm fh,min}\approx 1.1\,{\rm ks}$ 
in scenarios (i) and (ii) ---
intriguingly close to the observed time-scale.
Note that the fire hose instability only occurs
for $P_\parallel' > P_\perp'$;
in the reverse situation, a mirror instability exists for
$P_\perp'-P_\parallel' > 8.9B'^2/\mu_0$,
with growth time $\sim T_{\rm fh}'$.
\citet{gal94} argued that $P_\parallel'/P_\perp'$
increases from zero to unity downstream from the shock ---
the adiabatic index 
$(3+P_\parallel'/P_\perp')/(2+P_\parallel'/P_\perp')$
decreases from 3/2 to 4/3 as pitch-angle scattering
isotropizes the electrons ---
but $P_\parallel' > P_\perp'$ cannot be excluded.

Despite appearances, it is unlikely that
the fire hose instability causes the
unsteady, serpentine motions observed in the Vela X-ray jet
on time-scales between one day and several weeks
\citep{pav03},
because the growth time appears to be too short.
We estimate 
$T_{\rm fh,min}=
 0.14 \gamma_{\rm i} m_{\rm e} \langle\gamma\rangle/eB
 \approx 0.1 (\sigma/10^{-3})^{1/2}\,{\rm ks}$,
taking
$\langle\gamma\rangle=7\times 10^4$
[pair multiplicity $\approx 10^3$; see Figure 17 of
\citet{hib01}]
and
$B=0.15\sigma^{1/2}\,{\rm mG}$ (radial or toroidal),
where $\sigma$ is the ratio of Poynting to kinetic energy flux
at the base of the jet.
\citet{pav03} suggests an alternative scenario,
in which the end of the jet is bent by an external wind 
while the knots in the body of the jet are produced by
hydromagnetic (kink and sausage) instabilities on the local
Alfv\'{e}n time-scale.

Our time-lapse observations resolve the light-crossing 
time-scale of the smallest features in the field,
e.g.\ $2.2$ days for $0.\arcsec19$ at $2.0\,{\rm kpc}$.
Therefore, if the kilosecond variability we observe is real,
it must arise from
(i) a pattern traveling at a superluminal phase speed,
or (ii) relativistic Doppler boosting in the
upstream collimated outflow
($\gamma_{\rm i}^{-2}$ times the light-crossing time-scale;
cf.\ millisecond variability of unresolved gamma-ray bursters).
Our results are consistent with previous observations that 
also detected significant variations on, or faster than, 
the light-crossing time-scale:
\citet{hes02} observed the optical knots in the Crab PWN
to vary over six days, 
and \citet{pav03} observed the X-ray jet in the Vela PWN 
to vary over just two days.

\subsection{Radiation mechanisms
 \label{sec:gem5b}}
The near-infrared spectral indices displayed
in Table \ref{tab:gem1} are curious in several respects.
First, the inner knot has a steeper spectrum than
every other feature in the region --- not just in our data,
where the $J$-band flux is uncertain to $\pm 50$ per cent,
but also in data obtained with the Infrared Spectrometer
And Array Camera on the VLT
in $0.\arcsec65$--$0.\arcsec88$ natural seeing \citep{sol03}.
One explanation is that the inner knot is physically different
from the other features:
the wisps and sprite are part of the termination shock
and emit synchrotron radiation,
e.g.\ from ion-cyclotron-heated electrons
\citep{gal94},
whereas the inner knot lies upstream in the unshocked pulsar wind
and emits synchro-Compton radiation,
e.g.\ from electrons heated by magnetic reconnection 
\citep{cor90,lyu01a}
or parametric instabilities
\citep{mel96,mel98,mel02}
in a wave-like wind.
This explanation conflicts with recent simulations
which suggest that the inner knot is synchrotron emission
from an arch shock between the polar outflow and
equatorial backflow in the nebula
\citep{kom03}.
Unfortunately, we cannot discriminate between these two possibilities
spectrally, because both synchrotron and synchro-Compton radiation 
yield $\alpha=(1-p)/2$
given a power-law electron distribution
$N(\gamma) \propto \gamma^{-p}$
\citep{bla72,leu82}.
[Monoenergetic electrons in a large-amplitude wave
emit an inverse Compton spectrum
$F_\nu \propto \nu$ at frequencies below
$0.1\gamma^2 a^3 \Omega \sim 10^{16}\,{\rm Hz}$,
where $a=1$--10 is the wave nonlinearity parameter 
\citep{mel96,mel98} 
and $\Omega$ is the
pulsar spin frequency, but this low-frequency tail is modified to
$F_\nu \propto \nu^{(1-p)/2}$ for power-law electrons.]
Instead, we propose that the near-infrared polarizations of the 
inner knot and the sprite be measured.
Theory predicts that synchrotron and synchro-Compton
radiation are polarized perpendicular and parallel
to the magnetic field respectively;
one expects 50--80 per cent linear polarization from a linearly
polarized large-amplitude wave if $0.5\leq p\leq 5$
\citep{bla72}.
Therefore, assuming that the magnetic geometry of the wave-like wind
is similar at the inner knot and sprite
(e.g.\ helical at high latitude),
the polarization vectors of the two features are expected to be
perpendicular if the inner knot originates from the unshocked
pulsar wind and parallel if it is an arch shock.
Moreover, if the inner knot originates from
the unshocked wind, it should also be circularly polarized
(degree $\sim a^{-1}$, independent of $\nu$).

A second curious property of the spectra in Table \ref{tab:gem1}
is that the bright and faint wisps are steeper than the
sprite and rod. In this case, there is no doubt that both sets 
of features are shock-related synchrotron emission
at equatorial and polar latitudes respectively,
whether the sprite is located
at the working surface of the polar jet \citep{hes02} 
or along the arch shock \citep{kom03,del04}.
So why do the spectra differ?
One possibility is that the polarization of the large-amplitude wave
in the wind zone,
which is linear at the equator and circular at the pole,
affects the acceleration physics in the shock ponderomotively;
tentative indications to this end are emerging from recent
particle-in-cell simulations
(O.\ Skjaeraasen, private communication).

Table \ref{tab:gem1} raises a third puzzle: 
the near-infrared spectra of some features do not extrapolate smoothly 
to optical and X-ray wavelengths, while others do.
For example, the {\em HST} $V$-band surface brightness of the bright wisp
and the $V$-band flux of the sprite 
are measured to be $49\,\mu{\rm Jy\,arcsec^{-1}}$ and 
$4.1\,\mu{\rm Jy}$ respectively
\citep{hes95};
the same quantities, extrapolated from Table \ref{tab:gem1},
are predicted to be
$\approx 42\,\mu{\rm Jy\,arcsec^{-1}}$ and 
$\approx 15\,\mu{\rm Jy}$ respectively.
VLT and {\em HST} observations show that
the spectrum of the inner knot extends smoothly from near-infrared to
optical wavelengths, with $\alpha=-0.75\pm0.15$
\citep{sol03},
and this is corroborated (within larger uncertainties) by our data.
On the other hand, extrapolation of the VLT spectrum of the inner knot
to $0.16$--$0.32\,\mu{\rm m}$ yields
$F_\nu = 0.35\pm 0.05\,{\rm mJy}$,
significantly greater than the ultraviolet flux measured in \S\ref{sec:gem4b}.

The spectrum of the bright and faint wisps
is significantly steeper in X-rays than in the near infrared,
with $\alpha_{\rm X} \approx -1.3$ \citep{wei00}
and $\alpha_{\rm NIR} - \alpha_{\rm X} \approx 0.7$,
comparable to the steepening expected from synchrotron cooling
(Y.\ Lyubarsky, private communication).
The synchrotron cooling time for near-infrared-emitting electrons,
$t_{\rm cool}=2.8\times 10^4 (B/16\,\mu{\rm G})^{-3/2} 
 (\nu/2.4\times 10^{14}\,{\rm Hz})^{-1/2}\,{\rm yr}$,
is much longer than the flow time $t_{\rm flow}$ across the wisps.
On the other hand, the near-infrared spectral index of the sprite and rod
($\alpha = -0.21\pm 0.13$) is significantly shallower than the wisps
and similar to the average {\em radio} spectral index of the nebula
(Y.\ Lyubarsky, private communication).
This is either coincidental or highly surprising.
The sprite and rod are shock features in which one has
$t_{\rm flow} \ll t_{\rm cool}$ for near-infrared-emitting electrons
and kilosecond variability is observed;
they reflect electron acceleration at the present time.
The radio electrons
reflect the history of electron acceleration over the lifetime of the nebula 
and should be unaffected by the present dynamics of the sprite.
This paradox is encountered in a related context:
\cite{bie01} observed that the radio and optical wisps travel radially
in concert and display coordinated spectral index variations.

We conclude by reiterating that the measurement uncertainties
in our observations are substantial, due to the nonuniformity
of the nebula and PSF. Consequently, the evidence for
kilosecond variability and spectral differences,
while tantalizing, is inconclusive.
Improved observations --- preferably by an independent party
using a different instrument --- are essential
to clarify the situation.

\acknowledgments
We thank D. Barnes, A. Karick, M. O'Dowd, and J. Stevens 
for assisting with data reduction
and A. Oshlack for advice regarding error estimation.
This research was supported in part by 
Australian Research Council grant DP 0208735 (AM) 
and
NASA grant HST-AR-09215 (RWR).

This work is based on observations obtained at the 
Gemini Observatory, which is operated by the 
Association of Universities for Research in Astronomy, Inc., 
under a cooperative agreement with the NSF on behalf of the 
Gemini partnership: the National Science Foundation (United States), 
the Particle Physics and Astronomy Research Council (United Kingdom), 
the National Research Council (Canada), CONICYT (Chile), 
the Australian Research Council (Australia), CNPq (Brazil) 
and CONICET (Argentina).
Our observations were obtained with the Adaptive 
Optics System Hokupa'a/QUIRC, developed and operated by the 
University of Hawaii Adaptive Optics Group, with support from the 
National Science Foundation.

\bibliographystyle{apj}
\bibliography{psrwind}

\clearpage


\clearpage 

\begin{table}
\begin{tabular}{lcccc} \hline
Point source  &$J$ flux    &$K'$ flux    &$\alpha$   &length\\
              &[mJy]       &[mJy]        &           &[arcsec]\\
\hline
Star I        &6.25 (0.12) &3.09 (0.07)  &$ 1.24$   &---\\
Star II       &1.01 (0.02) &0.76 (0.02)  &$ 0.50$   &---\\
Pulsar        &3.21 (0.15) &2.62 (0.17)  &$ 0.36$   &---\\
Knot          &0.32 (0.16) &0.49 (0.17)  &$-0.78$   &---\\
\hline
Extended feature &$J$ flux    &$K'$ flux    &$\alpha$   &length\\
              &[$\mu$Jy arcsec$^{-1}$] &[$\mu$Jy arcsec$^{-1}$] & &[arcsec]\\
\hline
Bright wisp   &69.4 (1.4)  &97.7 (2.2)   &$-0.60$   &7.3\\
Faint wisp    &59.8 (1.3)  &80.3 (1.8)   &$-0.52$   &3.3\\
Rod           &62.8 (1.3)  &68.7 (1.6)   &$-0.16$   &1.2\\
Sprite        &46.2 (1.0)  &53.5 (1.2)   &$-0.26$   &2.5\\
\hline
\end{tabular}
\caption{De-reddened fluxes and power-law spectral indices 
($F_\nu \propto \nu^\alpha$) 
for features present in the images, 
collected into point sources (including the inner knot; top) 
and extended features (bottom). 
Fluxes for the extended features are quoted per arcsecond of length;
the lengths are shown in the right-hand column.  
Errors ($1\sigma$) are quoted in parentheses after the fluxes. 
A correction has been made for interstellar extinction (Eikenberry et al.\ 1997).
\label{tab:gem1}}
\end{table}




\end{document}